\documentclass[11pt,preprint]{aastex}

\newcommand{\kms}{\,km~s$^{-1}$}

\def\spose#1{\hbox to 0pt{#1\hss}}
\def\simlt{\mathrel{\spose{\lower 3pt\hbox{$\mathchar"218$}}
     \raise 2.0pt\hbox{$\mathchar"13C$}}}
\def\simgt{\mathrel{\spose{\lower 3pt\hbox{$\mathchar"218$}}
     \raise 2.0pt\hbox{$\mathchar"13E$}}}

\slugcomment{Accepted to AJ, Jan 2003 issue}

\shorttitle{Quasars Behind the Magellanic Clouds}
\shortauthors{Geha et al.}

\begin{document}


\title{Variability-Selected Quasars in MACHO Project Magellanic Cloud Fields}


\author{
      M.~Geha\altaffilmark{1},
      C.~Alcock\altaffilmark{2,3,4},
    R.A.~Allsman\altaffilmark{5},
    D.R.~Alves\altaffilmark{6},
    T.S.~Axelrod\altaffilmark{7},
    A.C.~Becker\altaffilmark{8},
    D.P.~Bennett\altaffilmark{3,9},
    K.H.~Cook\altaffilmark{3,4},
    A.J.~Drake\altaffilmark{3,10},
    K.C.~Freeman\altaffilmark{11},
      K.~Griest\altaffilmark{12},
    S.C.~Keller\altaffilmark{3}  
    M.J.~Lehner\altaffilmark{2},
    S.L.~Marshall\altaffilmark{3},
      D.~Minniti\altaffilmark{10},
    C.A.~Nelson\altaffilmark{3,13},
    B.A.~Peterson\altaffilmark{11},
      P.~Popowski\altaffilmark{14},
    M.R.~Pratt\altaffilmark{15},
    P.J.~Quinn\altaffilmark{16},
    C.W.~Stubbs\altaffilmark{4,15},
      W.~Sutherland\altaffilmark{17},
    A.B.~Tomaney\altaffilmark{15},
      T.~Vandehei\altaffilmark{12},
     D.L.~Welch\altaffilmark{18}\\
      {\bf (The MACHO Collaboration)}
      }

\altaffiltext{1}{UCO/Lick Observatory, University of California,
    Santa Cruz, 1156 High Street, Santa Cruz, CA 95064\\
    Email: {\tt mgeha@ucolick.org}}

\altaffiltext{2}{Department of Physics and Astronomy, University of
    Pennsylvania, Philadelphia, PA, 19104-6396\\
    Email: {\tt alcock, mlehner@hep.upenn.edu}}

\altaffiltext{3}{Lawrence Livermore National Laboratory, Livermore, CA 94550\\
    Email: {\tt kcook, adrake, skeller, cnelson, stuart@igpp.ucllnl.org}}

\altaffiltext{4}{Center for Particle Astrophysics, University of California,
    Berkeley, CA 94720}

\altaffiltext{5}{NOAO, 950 North Cherry Ave., Tucson, AZ 85719\\
    Email: {\tt robyn@noao.edu}}

\altaffiltext{6}{Columbia Astrophysics Laboratory,
     MailCode 5247, 550 W. 120th St., NY, NY, 10027\\
     Email: {\tt alves@astro.columbia.edu}}

\altaffiltext{7}{Steward Observatory, University of Arizona, Tucson, AZ 85721\\
    Email: {\tt taxelrod@as.arizona.edu}}

\altaffiltext{8}{Bell Laboratories, Lucent Technologies, 600 Mountain Avenue, 
    Murray Hill, NJ 07974\\
    Email: {\tt acbecker@physics.bell-labs.com}}

\altaffiltext{9}{Department of Physics, University of Notre Dame, IN 46556\\
    Email: {\tt bennett@bustard.phys.nd.edu}}

\altaffiltext{10}{Depto. de Astronomia, P. Universidad Catolica, Casilla 306,
    Santiago 22, Chile\\
    Email: {\tt dante@astro.puc.cl}}

\altaffiltext{11}{Research School of Astronomy and Astrophysics,
    Mount Stromlo Observatory, Cotter Road, Weston, ACT 2611, Australia\\
    Email: {\tt kcf, peterson@mso.anu.edu.au}}

\altaffiltext{12}{Department of Physics, University of California,
    San Diego, CA 92093\\
    Email: {\tt kgriest@ucsd.edu, vandehei@astrophys.ucsd.edu }}

\altaffiltext{13}{Department of Physics, University of California, Berkeley,
    CA 94720}

\altaffiltext{14}{Max-Planck-Institute f\"{u}r Astrophysik,
    Karl-Schwarzschild-Str.\ 1, 85748 Garching bei M\"{u}nchen, Germany \\
    E-mail: {\tt popowski@mpa-garching.mpg.de}}

\altaffiltext{15}{Departments of Astronomy and Physics,
    University of Washington, Seattle, WA 98195\\
    Email: {\tt stubbs@astro.washington.edu}}

\altaffiltext{16}{European Southern Observatory, Karl-Schwarzchild-Str.\ 2,
     85748 Garching bei M\"{u}nchen, Germany\\   
    Email: {\tt pjq@eso.org}}

\altaffiltext{17}{Department of Physics, University of Oxford,
    Oxford OX1 3RH, U.K.\\
    Email: {\tt w.sutherland@physics.ox.ac.uk}}

\altaffiltext{18}{Department of Physics and Astronomy, McMaster University,
    Hamilton, Ontario, Canada, L8S 4M1 \\
    Email: {\tt welch@physics.mcmaster.ca}}


\begin{abstract}

We present 47 spectroscopically-confirmed quasars discovered behind
the Magellanic Clouds identified via photometric variability in the
MACHO database.  Thirty-eight quasars lie behind the Large Magellanic
Cloud and nine behind the Small Magellanic Cloud, more than tripling the
number of quasars previously known in this region.  The quasars cover
the redshift interval $0.2 < z < 2.8$ and apparent mean magnitudes
$16.6\le \overline{V} \le 20.1$.  We discuss the details of quasar
candidate selection based on time variability in the MACHO database
and present results of spectroscopic follow-up observations.  Our
follow-up detection efficiency was 20\%; the primary contaminants were
emission-line Be stars in Magellanic Clouds.  For the 47 quasars
discovered behind the Magellanic Clouds plus an additional 12 objects
previously identified in this region, we present 7.5-year MACHO $V$-
and $R$-band lightcurves with average sampling times of 2-10 days.

\end{abstract}




\section{Introduction}
Techniques to find quasars, largely successful in other regions of the
sky, have had limited results towards the Magellanic Clouds
\citep{tin99,sch99,dob02}.  Crowding, recent star formation and
significant dust extinction cause major quasar surveys to avoid these
regions entirely, resulting in very few quasars known over the
substantial sky coverage of the Magellanic Clouds.  The most
successful selection method to date in this region has been at X-ray
wavelengths.  Although many tens of sources background to the
Magellanic Clouds have been identified in the X-rays
\citep{kah99,hab99,sas00}, counterparts to these sources at other
wavelengths have been stymied by postitional uncertainties; targeted
X-ray follow-up has allowed optical identification of $\sim 20$
extragalactic sources in the Magellanic Cloud region
\citep{cra97,sch99}.  The MACHO lightcurve database
\citep{alc97,alc00} provides an opportunity to search for quasars
behind the Magellanic Clouds via an alternative method: optical
variability.

Optical variability has been studied by many groups as a means of
constraining models of the quasar central engine \citep[][and
references therein]{hoo94,cri97,sir98, haw02}, as well as a method of
quasar identification behind globular clusters \citep{meu02}.
Although a handful of gravitationally-lensed quasars have well-sampled
lightcurves on the timescale of years \citep{alc02,hjo02}, most
studies have had short time baselines and poor resolution.  In one of
the largest optical monitoring programs, \citet{giv99} observed a
sample of 42 quasars over 7~years with an average sampling interval of
40~days.  Long-term optical variability of quasars in this study show
no strong evidence for underlying periodic structure.  This is in
sharp contrast to the majority of stellar sources in the Magellanic
Clouds which either do not vary or do so periodically.  We have used
this difference to separate quasar candidates from the overwhelming
stellar background in the Clouds.

A comprehensive search for quasars behind both the Large and Small
Magellanic Clouds (LMC and SMC, respectively) is motivated in part by
the lack of a suitable reference frame against which to measure the
proper motion of the Clouds.  Previous proper motion estimates have
suffered from an insufficient number or poorly distributed set of
reference objects \citep{jon94,kro97,ang00}.  Since the proper motion
of the Clouds is expected to be only a few mas/year, a well
distributed set of point-like background quasars could significantly
improve the accuracy of this measurement, constraining the orbital
history of these galaxies.  These objects may also prove useful as
light beacons for absorption line studies of the interstellar medium
in the Magellanic Clouds \citep{gib00,pro02}, as has been done for
suspected extragalactic X-ray sources in this region
\citep{kah01,hab01}.  Finally, this search was also motivated by
interest in the quasars themselves, in hope that the dense time
sampling of the MACHO lightcurves will provide clues to the physical
mechanisms underlying quasar light variation.

We discuss the MACHO database and our optical variability quasar
candidate selection methods in \S\,\ref{select}.  In
\S\,\ref{spectra}, we describe spectroscopic follow-up observations
and present 47 quasars discovered behind the Magellanic Clouds.  In
\S\,\ref{lightcurves}, MACHO lightcurves are presented for these
quasars.  Finally, in \S\,\ref{summary} we summarize our results and
discuss future quasar searches in this region.  Finding charts, light
curves and spectra for the quasars in this paper are available on
request from the authors or on a website given at the end of
\S\,\ref{summary}.

\section{Quasar Candidate Selection} \label{select}
\subsection{The MACHO Database}\label{macho}

The MACHO project monitored the Magellanic Clouds for the purpose of
detecting microlensing events between 1992 July and 2000 January
\citep{alc97, alc00}.  The Mount Stromlo Observatory 1.27-meter
telescope system provided simultaneous imaging in a red
($\rm\lambda\lambda5900$--$7800\mbox{\AA}$) and blue
($\rm\lambda\lambda4370$--$5900\mbox{\AA}$) filter, over a 0.5
$\Box^{\circ}$ field of view with a scale of $0.63''$ per pixel.  We
monitored 82 fields (35 $\Box^{\circ}$) in the LMC, and 6 fields (2.5
$\Box^{\circ}$) in the SMC; the location of these fields is shown in
Figures~\ref{lmc} and \ref{smc}.  Average sampling frequencies varied
from 2 to 10 days between fields.  Exposure times were 300-s and 600-s
for the LMC and SMC, respectively.  Photometric transformations from
MACHO passbands to standard $V$- and $R$-band magnitudes proceeded
using the calibrations discussed in \citet{alc99}.  Since the MACHO
passbands are non-standard and these photometric transformation were
determined for stars and not for emission-line quasars, the $(V-R)$
quasar colors in this paper should be approached with caution.  We
consider only measurements with photometric errors less than 5\%; the
average MACHO quasar lightcurve contains 600 good photometric
measurements over 7.5 years.

Selection of the quasar candidates, described below, was performed on
the first 5.7~years of MACHO data; the final 7.5-year MACHO light
curves are presented in the lightcurve analysis of
\S\,\ref{lightcurves}.  Due to limited data access at the time of
selection in the outer LMC MACHO fields, the variability search
discussed in \S\,\ref{vselect} was spatially restricted to the SMC and
30 MACHO LMC fields.  In \S\,\ref{xselect}, the full spatial coverage
of the MACHO database was searched for variable counterparts to known
radio and X-ray sources.  The final MACHO Magellanic Cloud database,
37.5$\Box^{\circ}$ over 7.5~years, has recently become readily
accessible and variability selection will be run on the outer LMC fields
not analysed in this paper.

\subsection{Quasar Candidate Variability Selection}\label{vselect}

\begin{figure}
\epsscale{0.9}
\plotone{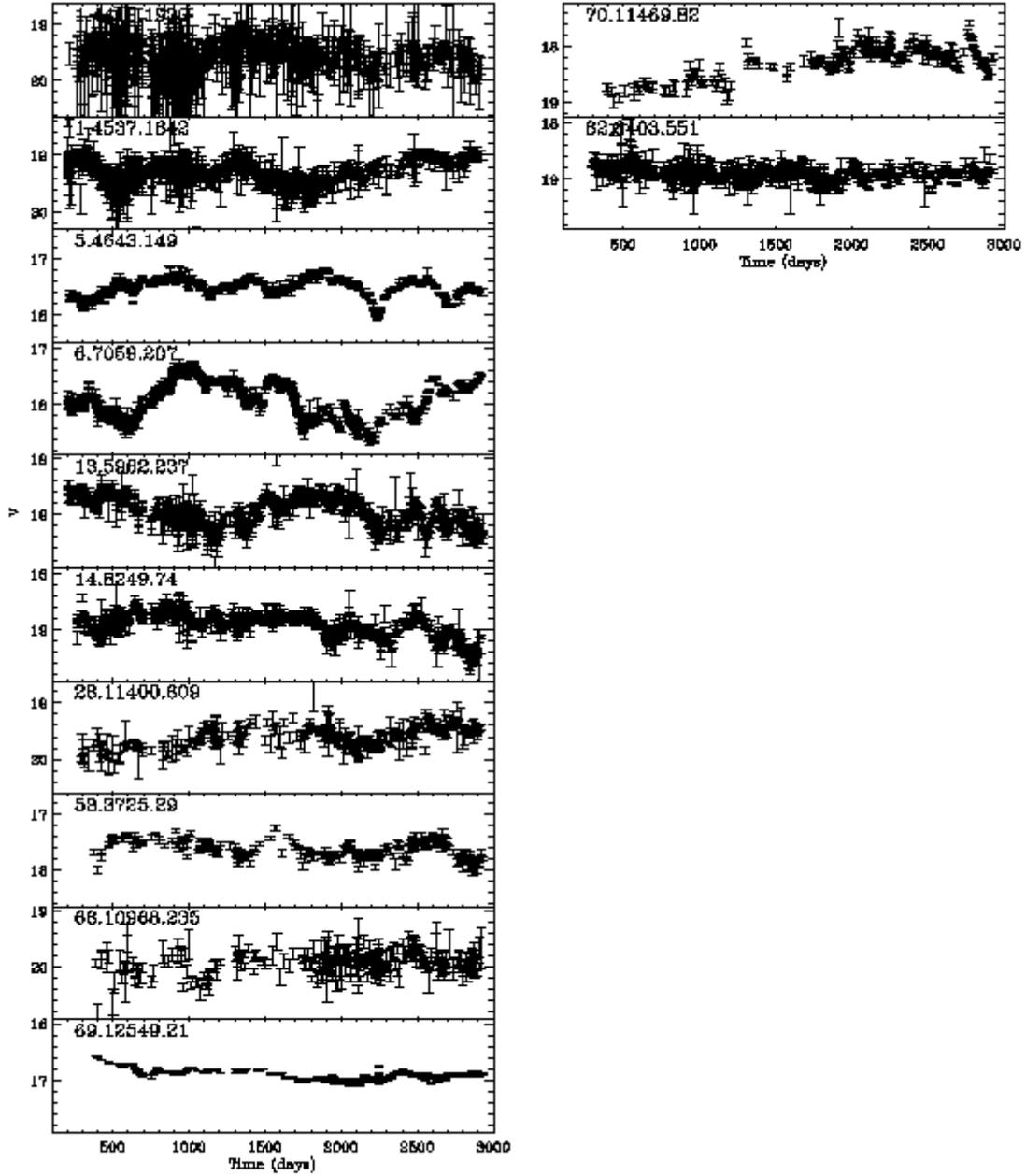}
\vskip 0.5 cm
\caption{$V$-band MACHO lightcurves for 12 quasar/AGNs in the MACHO
database previously known behind the LMC.  See Table~1 for reference to the
discovery paper of each source.
\label{lc_xray}}
\end{figure}

At the time of candidate selection, the long-term optical photometric
behavior of quasars was not adequately constrained to fully automate
quasar selection in the MACHO database.  Instead, the selection method
was designed to automatically reject known classes of variable stars,
with the final step being a selection by eye.  The MACHO database
contained 9 Active Galactic Nuclei (AGN) listed in Table~1 which had
been previously cross-identified with X-ray sources by \citet{sch99}
and one AGN which was serendipitously discovered by \citet{bla86}.
Two additional sources in the MACHO database were presented by
\citet{dob02} subsequent to candidate selection.  Lightcurves for
these sources, shown in Figure~\ref{lc_xray}, provided a training set
around which the selection method was developed.  The final subjective
step was to select, by eye, lightcurves similar to these known
sources.  Since our goal was to identify a robust set of quasars,
rather than a complete census of quasars behind the Clouds, we deemed
this level of subjectivity to be acceptable.
  
\begin{figure}
\plotone{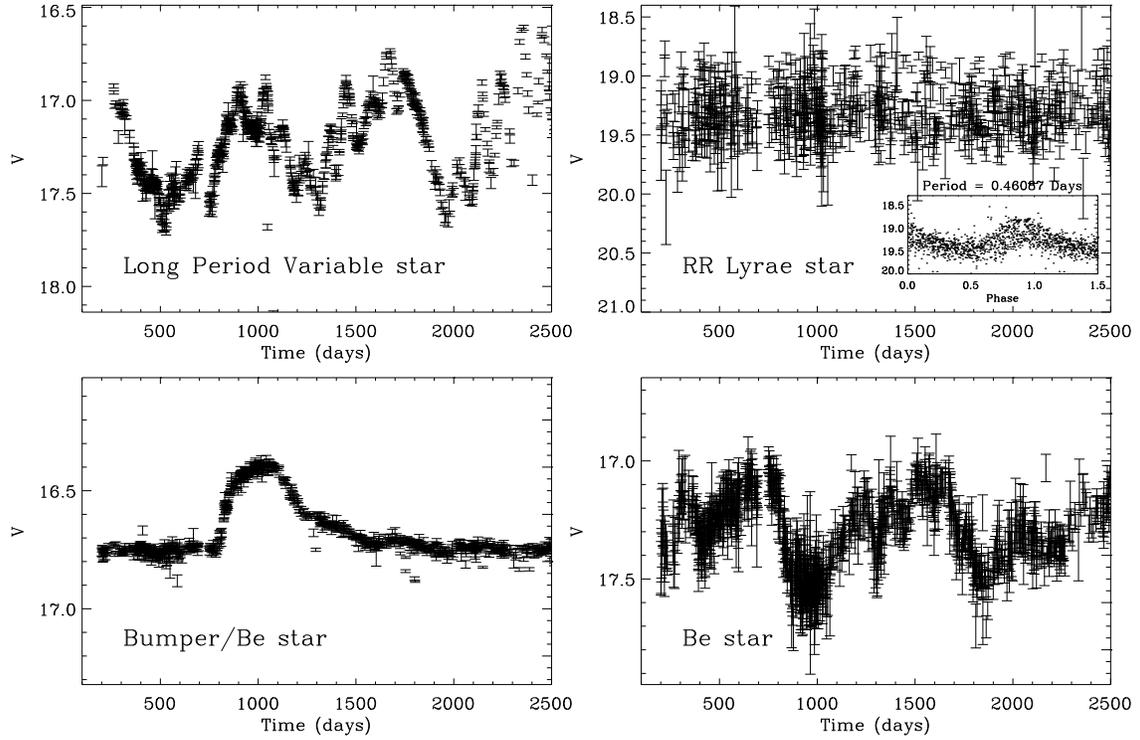}
\vskip -3 cm
\caption{$V$-band lightcurves of variable stars in the MACHO database.
Examples of stars rejected by our quasar selection technique are ({\it
top, left panel}) red, long period variable stars, ({\it top, right
panel}) RR Lyrae stars, in this example phased with a 0.46 day period
(panel inset), and ({\it bottom, left panel}) blue variable ('Bumper')
stars.  The main contaminant to our quasar candidate search were
emission line Be stars with quasi- to aperiodic lightcurves ({\it
bottom, right panel})\label{fig_lc_ex}.}
\end{figure}

Variability selection was run on 30 MACHO LMC fields
(15$\Box^{\circ}$) containing 12 million objects and 6 MACHO SMC
fields (2.5$\Box^{\circ}$) containing 2 million objects.  Candidate
selection began with the Level 1 MACHO database, a data subset
containing 140,000 objects flagged as having a significant deviation
from a constant brightness lightcurve \citep{alc00}.  We required
objects to have a minimum of 50 photometric measurements in
$V$- and $R$-bands.  Weighted average magnitudes were calculated from
the standard equation:
\begin{equation}\label{meanV}
\overline{V} \equiv \sum_{i=1}^{N} \frac{V_i}{\sigma_{V,i}^2}~ \Big{/}~ \sum_{i=1}^{N}\frac{1}{\sigma_{V,i}^2}.
\end{equation}
\noindent
where $N$ is the total number of individual photometric measurements
$V_i$ with associated errors $\sigma_{V,i}$.  We considered candidates
between $16 \le \overline{V}\le 20$ which is one magnitude
brighter (fainter) than the MACHO photometric completeness (saturation)
limits.  Candidates were required to have weighted average colors
bluer than $(\overline{V-R}) \le 1.0$.  This color cut eliminated
long-period quasi-periodic variable stars, while retaining all of the
training set AGNs.  An example lightcurve of a long period variable is
shown in the top right panel of Figure~\ref{fig_lc_ex}; the majority
of these stars are extremely red and otherwise difficult to remove
from the final quasar candidate list.

Two statistics were used to quantify the amount of lightcurve
variability.  First, the intrinsic variability for each candidate
quasar lightcurve was calculated as:
\begin{equation}\label{intr_var}
\widehat{\sigma_V} \equiv 
     \sqrt{ \frac{\sum \big{(}V_i - \overline{V}\big{)^2}}{N-1}
          - \frac{\sum \sigma_{V,i}^2}{N} }
\end{equation}
\noindent
This quantity is an estimate of the true source variability in the
absence of photometric errors.  The first term is the total variance
measured from the lightcurve data, while the second is an estimate of
the variance due to photometric measurement errors alone.  We note
that the photometric error estimate supplied by the photometry code
Sodophot \citep{alc97} is known to have some bias for bright stars.
This bias is on the order of 0.01 magnitudes and should have only a
small effect on the current data.  \citet{giv99} monitored a sample of
42 quasars monthly for 7~years, measuring intrinsic variabilities
between $0.05 \le \widehat{\sigma_B} \le 0.32$ in the $B$-band.  We
therefore require the intrinsic variability of MACHO candidate quasars
to be larger than 0.05 magnitudes ($\widehat{\sigma_{V}} \ge 0.05$).
The second variability statistic calculated for each lightcurve is the
variability index \citep{wel93}.  This index is a measure of
correlated variability between the MACHO $V$- and $R$-bands, defined
as:
\begin{equation}
I \equiv \sqrt{\frac{1}{N(N-1)}} ~ \sum_{i=1}^N 
                \Big{(}\frac{V_i - \overline{V}}{\sigma_{V,i}}\Big{)}~ 
                \Big{(}\frac{R_i - \overline{R}}{\sigma_{R,i}}\Big{)}
\end{equation}
This quantity approaches zero for uncorrelated variability.  Quasar
variability is expected to be highly correlated between the MACHO
passbands; we require $I\ge 1.5$ for our candidates.  This limit was
set slightly below the minimum $I$-value determined from the training
set AGN.  Due to MACHO noise characteristics the two variability
statistics described above are not redundant.  Systematic noise terms
in the MACHO system which affect a single filter can cause large
$\widehat{\sigma_V}$ values, but may be eliminated as this variability
is not correlated in the second filter.  Conversely, variable
observing conditions, such as seeing and sky brightness, can cause
correlated variability, but these measurements often have large
photometric errors and can be rejected based on our
$\widehat{\sigma_V}$ cut.
 
Periodic variable stars, such as Cepheid and RR Lyrae stars, were
removed using the MACHO Variable Star Catalogue.  This catalogue
contains period information for the majority of variable MACHO objects
as determined from a super-smoother algorithm which models folded
lightcurves \citep{coo95}.  We reject as candidate lightcurves for
which a period ($\tau$) was found at high significance in both
passbands over the range $0.1 \le \tau \le 500$ days.  This cut did
not include aliased frequencies ($\tau = 1/n$ days, $n$=1,2,..).  An
example lightcurve of a typical RR Lyrae star in the MACHO database
rejected by our periodicity cut is shown in the top right panel of
Figure~\ref{fig_lc_ex}.

In the final quasar candidate selection step, each candidate
lightcurve was examined by eye to remove objects with spurious noise
characteristics or quasi-periodic components.  Roughly 2500 light
curves were examined by eye; a total of 360 lightcurves were
considered candidate quasars.  A fraction of lightcurves rejected by
eye were due to noise effects above the thresholds set by our
variability cuts.  However, the majority of rejected candidates were
quasi-periodic lightcurves characteristic of blue variable stars,
known to the microlensing community as a 'Bumper' stars \citep{coo95}.
An example of such a lightcurve is shown in the bottom left panel of
Figure~\ref{fig_lc_ex}.  These stars typically have strong Balmer
emission lines at the velocity of the Clouds and are associated with
the Be star phenomena \citep{kel02}.  These blue variables often have
quasi-repeatable outbursts and can be eliminated from the candidate
list.  As designed, our quasar selection technique successfully
recovered all of the previously known quasar/AGNs in the search
region.

\subsection{Additional Candidate Selection Methods}\label{xselect}

We have additionally searched for optically-variable counterparts in
the error boxes of suspected extragalactic radio and X-ray sources in
several Magellanic Cloud surveys.  This allowed us to extend our
search to the full spatial coverage of the MACHO database
(37.5$\Box^{\circ}$), as explained in \S\,\ref{macho}.  In the LMC, we
have searched the LMC radio catalogues of \citet{mar97} and
\citet{fil98} and the X-ray catalogue of \citet{hab99}; in the SMC we
have searched the radio catalogue of \citet{fil97} and the X-ray
catalogues of \citet{kah99} and \citet{sas00}.  Photometric
constraints, described in the previous section, are relaxed in this
search: any aperiodic MACHO variable ($\widehat{\sigma_V} > 0$) inside
the $1\sigma$ spatial error box of a cataloged source (typically
$\sim15''$) is considered a candidate.  We identified variable,
aperiodic MACHO counterparts in the $1\sigma$ error boxes for $\sim
5\%$ of the cataloged objects.  In the last column of Tables 2 and 3,
we note which quasars were identified by this method.

\section{Spectroscopic Confirmation of 47 MACHO Quasars}
\label{spectra}

Follow-up spectroscopic observations for our quasar candidates were
obtained with the Anglo-Australian Telescope and 2dF multifiber
spectrograph (AAT+2dF) in 1999 October and 2001 January.  Additional
observations were made with the Australian National University 2.3m
and Double Beam Spectrograph (ANU+DBS) over 8 nights between 1999
March and 2000 November.  The AAT+2dF is a fiber-fed spectrograph
with $\sim2''$ diameter fibers and 200 fibers available per pointing.
These observations were made through the 300B grating covering
$\rm\lambda\lambda3800$--$7800\mbox{\AA}$ with $4.3\mbox{\AA}$ pixels.
The 2dF data were reduced using the standard 2dFdr reduction pipeline
software \citep{bai02}.  Observations with the ANU+DBS were made through
a single, $2''$ wide slit in combination with a low resolution
$2.2\rm\mbox{\AA}$ per pixel blue grating
($\rm\lambda\lambda3500$--$6000\mbox{\AA}$) and a red
$4.0\rm\mbox{\AA}$ per pixel grating
($\rm\lambda\lambda6000$--$9000\mbox{\AA}$).  These data were reduced
with IRAF single long-slit spectra reduction procedures.  A total of 259
candidate quasars were observed: spectra for 220 candidates were
obtained in six separate AAT+2dF pointings, 39 candidates were
observed in single pointing with the 2.3m+DBS.  The remaining $\sim
100$ candidates will be observed in future observing runs.

Spectroscopic follow-up revealed a total of 47 previously unknown
quasars behind the Magellanic Clouds: 38 behind the LMC and 9 quasars
behind the SMC.  The quasars cover the redshift interval $0.2 < z <
2.8$ and range in apparent magnitude between $16.6\le \overline{V} \le
20.1$ mag.  For the majority of objects, redshifts were determined
from two or more broad lines; rest-frame spectra for all 47 newly
discovered quasars are shown in Figure~\ref{fig_cspec}.  All
objects appear unresolved in MACHO images.  Absolute magnitudes were
computed assuming a consensus cosmology of $\Omega_m = 0.3$,
$\Omega_\Lambda = 0.7$ and H$\rm_o$ = 75\kms Mpc$^{-1}$ and are
corrected for reddening.  Values for reddening vary in the Magellanic
Clouds due to patchy dust distribution, and accurate estimates do not
yet exist in the directions of all of our quasars.  We instead
adopted, for each Cloud, a single reddening value of $E(B-V) = 0.12$
and $0.05$ \citep{dut01} for the LMC and SMC, respectively, and a
Galactic extinction law.

\begin{figure}
\epsscale{0.93}
\plotone{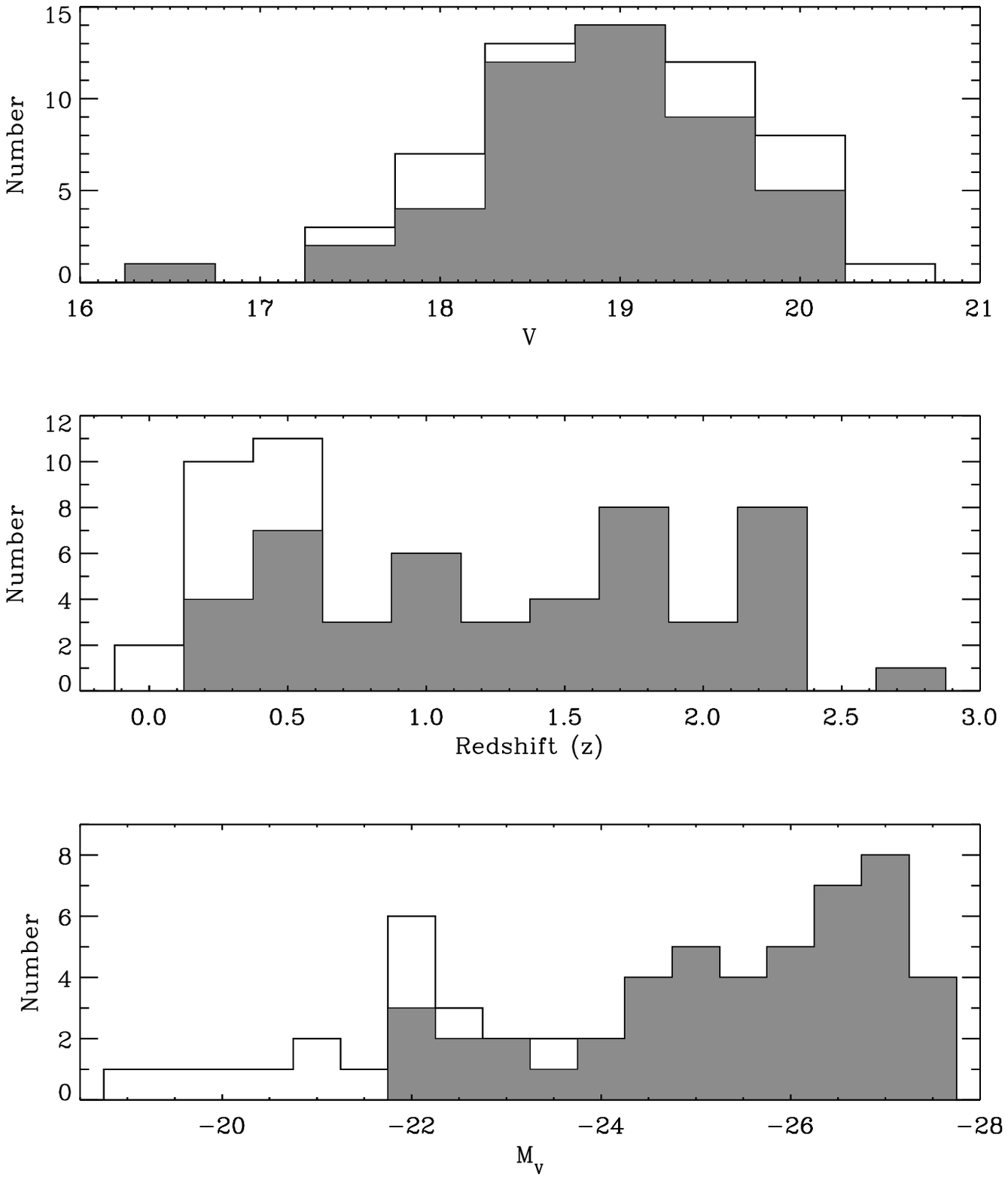}
\caption{Histograms of MACHO quasars as a function ({\it from top to
bottom}) of mean apparent $V$-band magnitude, redshift, $z$, and
absolute magnitude, $M_V$.  The shaded region of each histogram
represents quasars identified in this study behind the Magellanic
Clouds. \label{fig_hist}}
\end{figure}

The spatial distribution of the discovered MACHO quasars is shown
relative to the Magellanic Clouds in Figures~\ref{lmc} and \ref{smc}.
In the LMC, quasars appear concentrated to the South of the LMC bar
due to several 2dF spectroscopic pointings in this region.  Given that
$\sim100$ quasar candidates have not yet been spectroscopically
followed up in this region, and subjectivity in the candidate
selection process, our quasar sample is unlikely to be complete.  In
Tables~2 and 3 we list for each source a MACHO ID number, equatorial
coordinates $\alpha$(J2000) and $\delta$(J2000), weighted average
apparent $\overline{V}$ magnitude and $(\overline{V-R})$ color, the
redshift, z, the mean absolute magnitude, $M_V$, and the number of
$V$-band photometric measurements, $n_V$, available in the MACHO
database.  We also note if the object was selected as a candidate
based on optical variability inside the error box of a suspected
extragalactic radio and X-ray source.  Histograms of the quasars,
distributed as a function of redshift, apparent and absolute
magnitude, are shown in Figure~\ref{fig_hist}.

\subsection{Candidate List Contamination: Be/Ae Stars}

The primary contamination during spectroscopic follow-up of our quasar
candidates were emission-line main sequence stars (Be/Ae stars) in the
Magellanic Clouds.  These stars display strong Balmer
emission/absorption lines at the velocity of the Clouds and are known
to be photometrically variable \citep{hub98}.  Of the 258 quasar
candidates observed, 188 were emission-line Be/Ae stars.  The brightest ($V
\le 16.5$), more well studied, emission-line stars have a strong
periodic component to their lightcurves \citep[`Bumpers';][]{coo95}
and were already removed from the quasar candidate list.  However,
fainter Be/Ae stars appear to be quasi- to aperiodic variables and
were thus included as quasar candidates.  An example Bumper light
curve is compared to a fainter Be/Ae star in the bottom panels of
Figure~\ref{fig_lc_ex}.  These objects have very similar spectra,
despite significantly different lightcurve behavior.  A preliminary
analysis of this type of blue variable star is presented in \citet{kel02}.

\section{MACHO Quasar Lightcurves}\label{lightcurves} 

$V$-band lightcurves for all quasars listed in Tables 1, 2 and 3 are
presented in Figures~\ref{lc_xray}, \ref{lc}, and \ref{lc_smc}.  The
lightcurves have on average 600 photometric measurements in both $V$-
and $R$-band, spanning 7.5~years.  The zero point of the time axis of
these plots corresponds to JD 2448623.5, or UT 1992 January 2.0.  We
have searched for close quasar pair candidates in a $15''$ radius
around each source, but find no aperiodic lightcurves at these
distances.  We note that, unlike the previously discovered sources
listed in Table~1 which are predominately AGN, the majority of MACHO
sources would be classified as quasars based on the usual absolute
luminosity criteria of $M_B = -23$ (strict classification is prohibited
by the fact that we determine $M_V$ rather than $M_B$ magnitudes).

Similar to the results of \citet{giv99}, we find that the majority of
quasars become bluer in $(V-R)$ as they brighten.  To quantify this, we
calculate the Spearman's rank correlation probability ($P_r$) between
magnitude and color measurements for each source.  The correlation is
in the same sense for all quasars (bluer $V-R$ colors at brighter $V$
magnitudes), and is highly significant at a $4.5\sigma$ level or
greater ($P_r < 3 \times 10^{-5}$) for 95\% of the quasar lightcurves.
In contrast, the majority (63\%) of contaminating Be/Ae stars in our
spectroscopic sample became redder as they brightened at similar
significance levels.  This distinction will be used to improve future
quasar selection criteria.

\begin{figure}
\epsscale{1}
\plotone{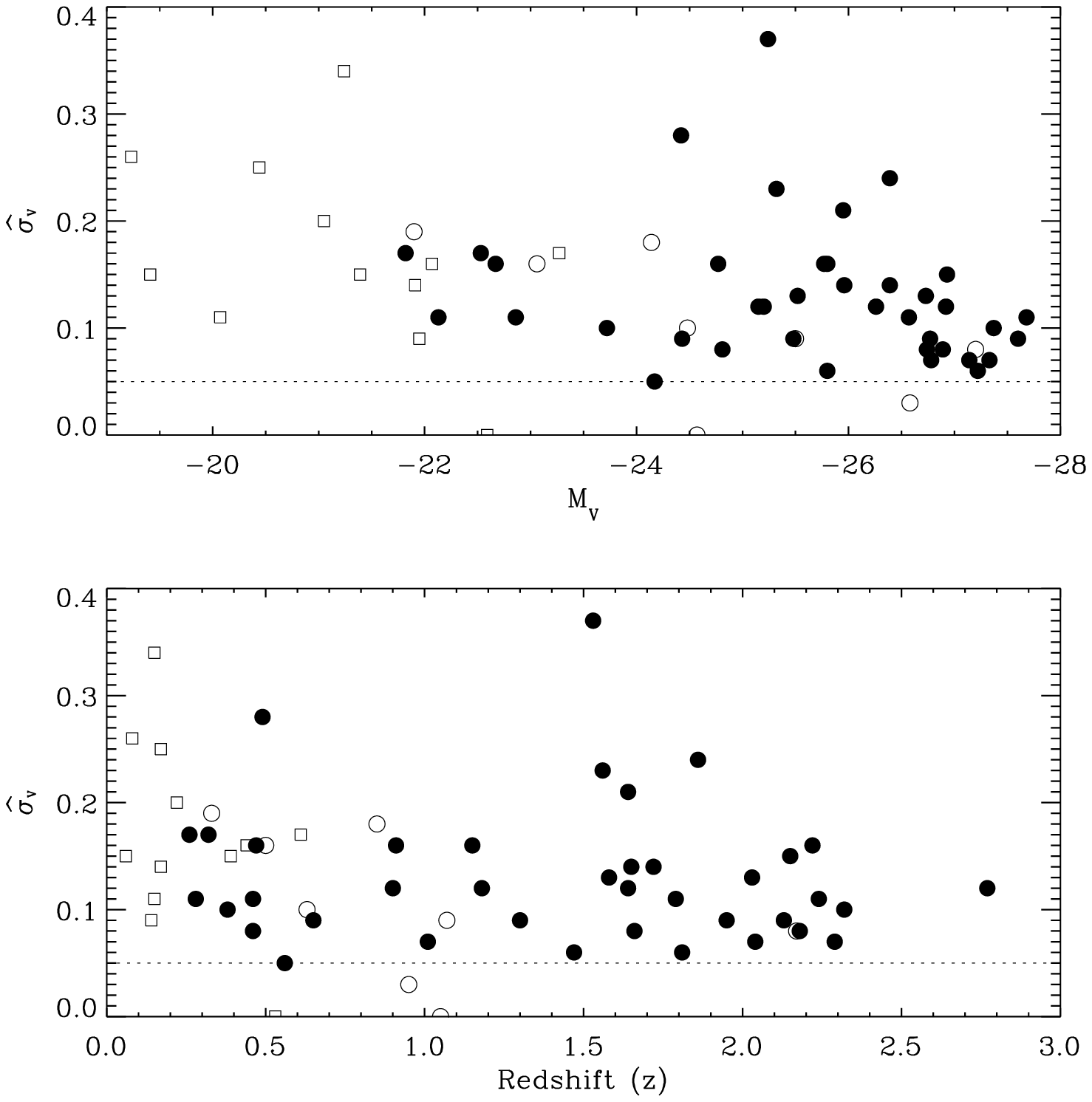} 
\caption{The intrinsic photometric variability $\widehat{\sigma_{V}}$
as a function of absolute magnitude, $M_V$ ({\it top panel}) and
redshift ({\it bottom panel}).  The dotted line indicates the
variability selection requirement of $\widehat{\sigma_{V}} \ge 0.05$.
Solid circles represent MACHO quasars discovered via variability-only
selection, open circles indicate quasars discovered as variable
counterparts to X-ray and radio sources for which the amount of
variability required for selection was relaxed.  Open squares
represent previously discovered quasar/AGNs.
\label{sig} }
\end{figure}

In Figure~\ref{sig}, the $V$-band intrinsic variability, as defined in
Eqn.~2, is plotted as a function of quasar absolute luminosity and
redshift.  For the more homogeneous subset of quasars discovered via
our variability-only method (Fig.~\ref{sig}, solid points), we find
according to the Spearman's rank test, a very weak correlation at the
$2.2\sigma$ level between variability and absolute luminosity
($\widehat{\sigma_V}$ - $M_V$) in the sense that intrinsically
luminous quasars tend to be less variable.  For the full set of
objects shown in Figure~\ref{sig}, the same correlation between
$\widehat{\sigma_V}$ - $M_V$ was more significant at the $3.3\sigma$
level.  No correlation was found between the intrinsic source
variability and redshift ($\widehat{\sigma_V}$ - $z$) for our quasar
subset.  For the full set, a weak negative correlation between
variability and redshift is detected at the $2.5\sigma$ level.
Several groups have claimed a similar correlation between variability
and absolute luminosity \citep{hoo94,cri97,giv99}, however, both
positive and negative correlations have been claimed between
variability and redshift.  These results are difficult to interpret
due to strong correlations between redshift and absolute luminosity,
as well as varying definitions of variability.  The quasar light
curves presented in this paper have sufficient time resolution to
allow power spectrum and other detailed lightcurve analyses.  It is
hoped that such work will be easier to interpret and have interesting
implications for the physical mechanisms driving the AGN/quasar
variability.

\section{Discussion and Summary}\label{summary}

We present 47 quasars discovered behind the Magellanic Clouds: 38
behind the LMC and 9 behind the SMC, significantly increasing the
number density of known quasars in this region.  The quasars cover the
redshift interval $0.2 < z < 2.8$ and apparent mean magnitudes
$16.6\le \overline{V} \le 20.1$.  Candidate quasars were identified
based on aperiodic variability in the MACHO database.  MACHO light
curves are presented for the newly discovered quasars as well as 12
quasar/AGNs identified in previous studies.  The primary contamination
during spectroscopic follow-up were quasi- or aperiodic Be/Ae stars in
the Magellanic Clouds.  Spectroscopic follow-up of quasar candidates
in the MACHO database is not yet complete.  In the outer LMC, 52 MACHO
fields have become accessible for variability-only (\S\,\ref{vselect})
selection since the original candidate selection was run.  In
addition, $\sim 100$ candidates have not yet been followed-up from the
search presented in this paper.  We therefore expect the MACHO
database to yield many more sources in the future.

A similar photometric variability selection has been applied to
another microlensing database in the Magellanic Clouds, OGLE-II
\citep{eye02}, but has not yet been followed-up spectroscopically.  We
have checked our sample of quasars against the OGLE candidate list
presented by \citeauthor{eye02}.  Four of the MACHO quasars are listed
as candidates in this paper, (OGLE candidates: L92, L114, L155, S12),
eight OGLE candidates are within $1''$ of spectroscopically confirmed
Be stars (L51, L87, L103, L121, L148, L153, S8, S25) and six OGLE
candidates are common to our list which have not yet been followed-up
spectroscopically.  Despite similar quasar candidate selection
strategies, the majority of the OGLE candidates did not make our
candidate list.  This is due in part to our final subjective
rejection of Be/Bumper lightcurves: 24 objects classified as quasar
candidates by \citeauthor{eye02} were considered Be/Bumper star
candidates according to MACHO photometry; the remaining objects in the
OGLE list either did not pass our minimum variability cut or fell
outside MACHO fields.  Continued follow-up of both MACHO and OGLE
candidates is certain to increase significantly the number quasars in
the Magellanic Cloud region.

Quasars behind the Magellanic Clouds are extremely useful tools with
which to study the Clouds themselves.  Our motivation for this study
was to uncover a robust set of reference objects against which to
measure the proper motion of the Magellanic Clouds.  First epoch {\it
Hubble Space Telescope} images have been scheduled for a subset of
quasars presented in this paper; second epoch imaging is expected to
allow an estimation of the Cloud's orbital motion with sufficient
accuracy to constraint models of the Galactic halo.  Quasars are also
invaluable tools for a variety of other studies, for example as probes
of the Clouds' interstellar medium.  We therefore provide finding
charts, lightcurves and spectra for all quasars presented in this
paper, available electronically at
http://www.ucolick.org/$\sim$mgeha/MACHO or on request from the
authors.

\acknowledgments

We would like to thank Terry Bridges for help with 2dF observations
and reductions. This work was performed under the auspices of the
U.S. Department of Energy, National Nuclear Security Administration by
the University of California, Lawrence Livermore National Laboratory
under contract No. W-7405-Eng-48.  DM is supported by FONDAP Center
for Astrophysics 15010003.  MG was supported in part by a grant from
IGPP-LLNL UCRP.




\clearpage

\begin{figure}
\epsscale{0.8}
\caption{$R$-band image of the Large Magellanic Cloud ($8^{\circ}
\times 8^{\circ}$; G. Bothun 1997, private communication) with the
MACHO photometric coverage indicated by large numbered squares.  North
is up, East is to th left in this image.  Small black and white
symbols indicate quasars discussed in this paper.  The distribution of
quasars is biased due to early unavailability of analyzed MACHO fields
and incomplete spectroscopic follow-up coverage.
\label{lmc}}
\end{figure}

\begin{figure}
\epsscale{0.55}
\caption{$R$-band image of the Small Magellanic Cloud ($3^{\circ}
\times 5^{\circ}$) with the MACHO photometric coverage
indicated. Small black and white symbols indicate presented quasars. \label{smc}}
\end{figure}

\clearpage

\begin{figure}
\epsscale{0.8}
\plotone{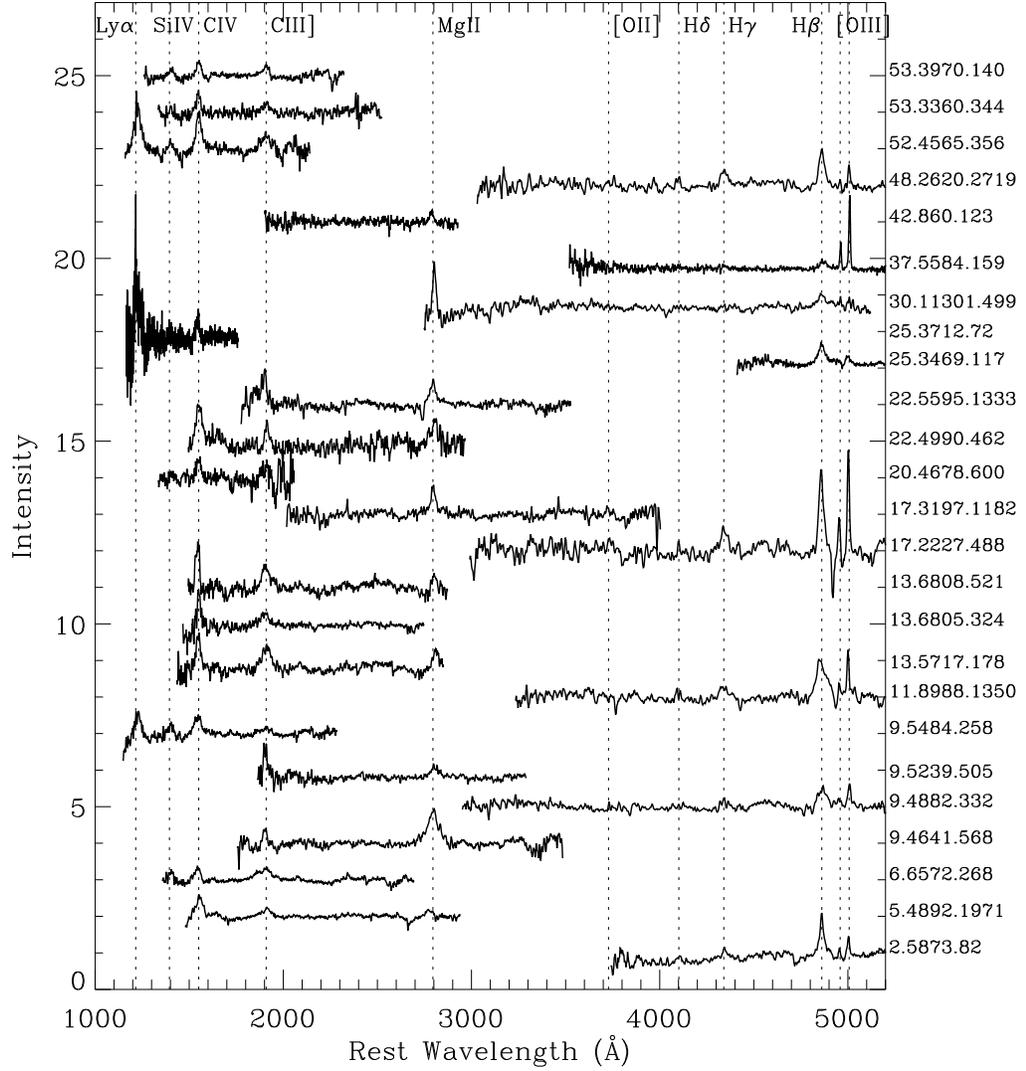}
\vskip -0.5 cm
\caption{Low resolution spectra, corrected to the rest frame, for the
47 MACHO quasars presented.  Source names appear to the right of each
spectrum.  The spectra have been normalized, smoothed and arbitrarily
shifted in intensity for display purposes.
\label{fig_cspec}}
\end{figure}

\begin{figure}
\epsscale{0.8}
\plotone{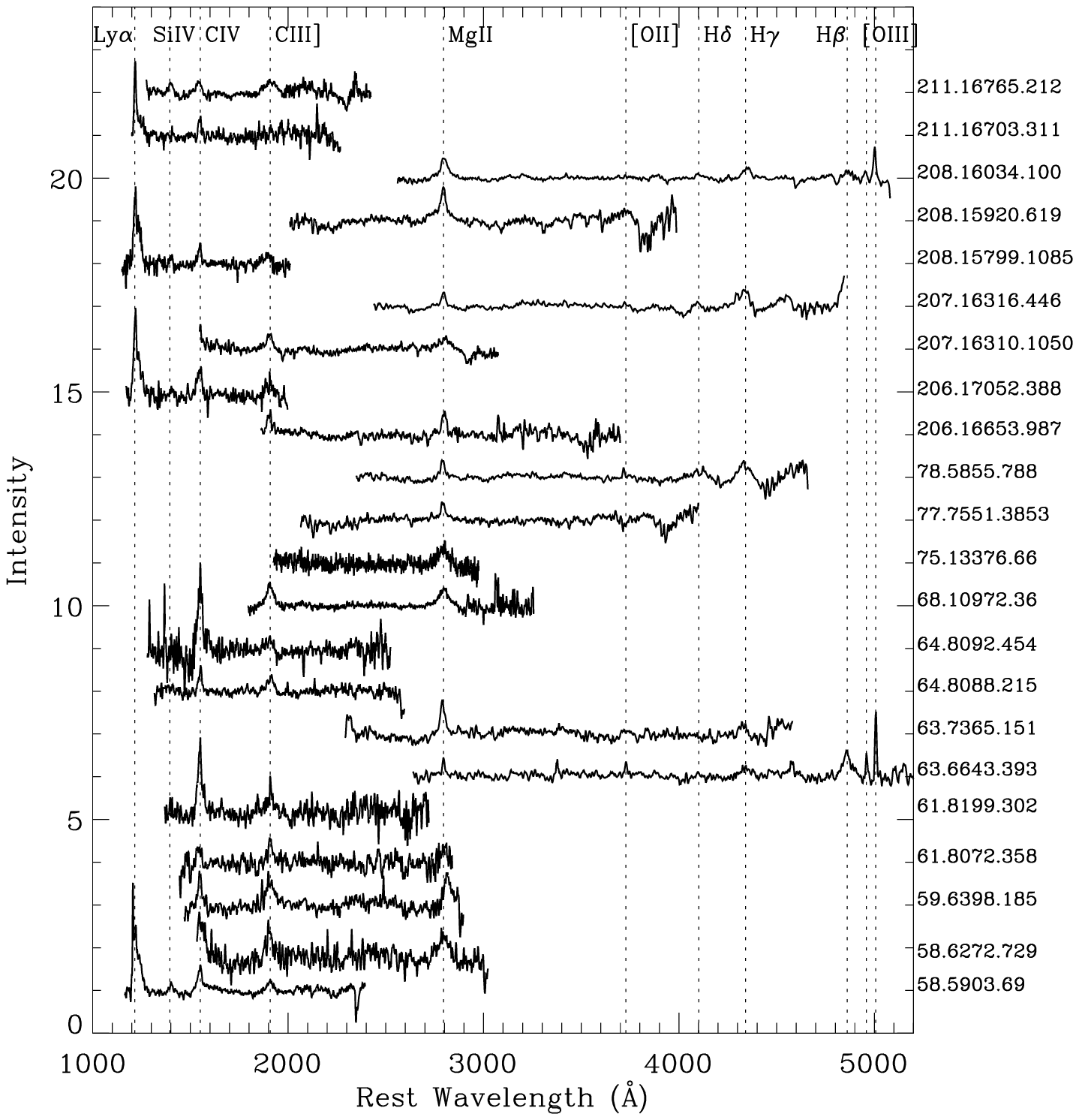}
\addtocounter{figure}{-1}
\vskip -0.5 cm
\caption{continued.}
\end{figure}

\begin{figure}
\epsscale{0.9}
\plotone{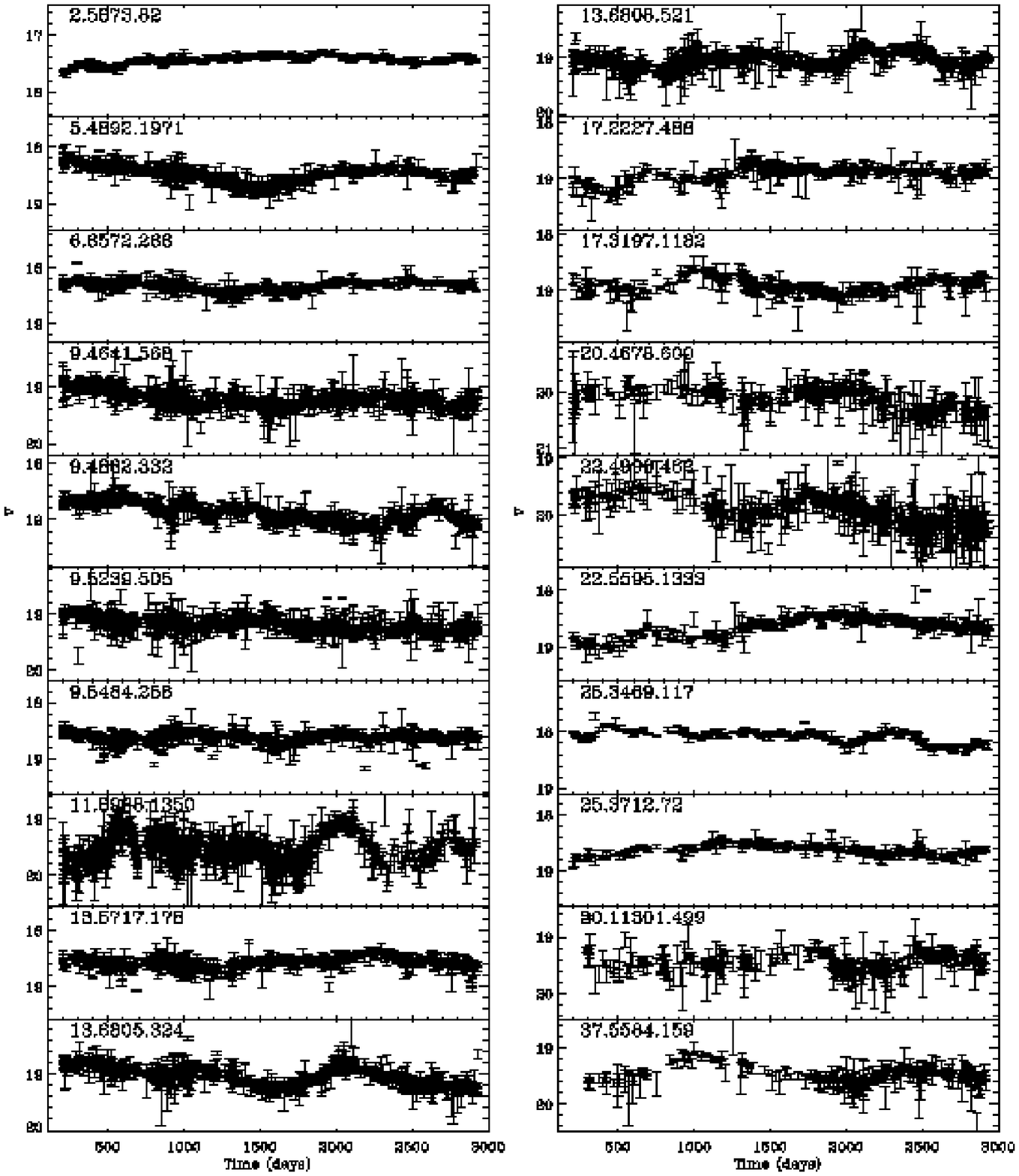}
\vskip 0.5 cm
\caption{$V$-band MACHO lightcurves for quasars behind the LMC
presented in this paper.  The y-axis scale is the same for all
objects, spanning 2.0 magnitudes centered on the weighted mean $V$-band
brightness of each source.  The lightcurves cover 7.5 years; the
zeropoint of the times axis corresponds to JD 2448623.5, or UT 1992
January 2.0.
\label{lc}}
\end{figure}

\begin{figure}
\epsscale{0.9}
\plotone{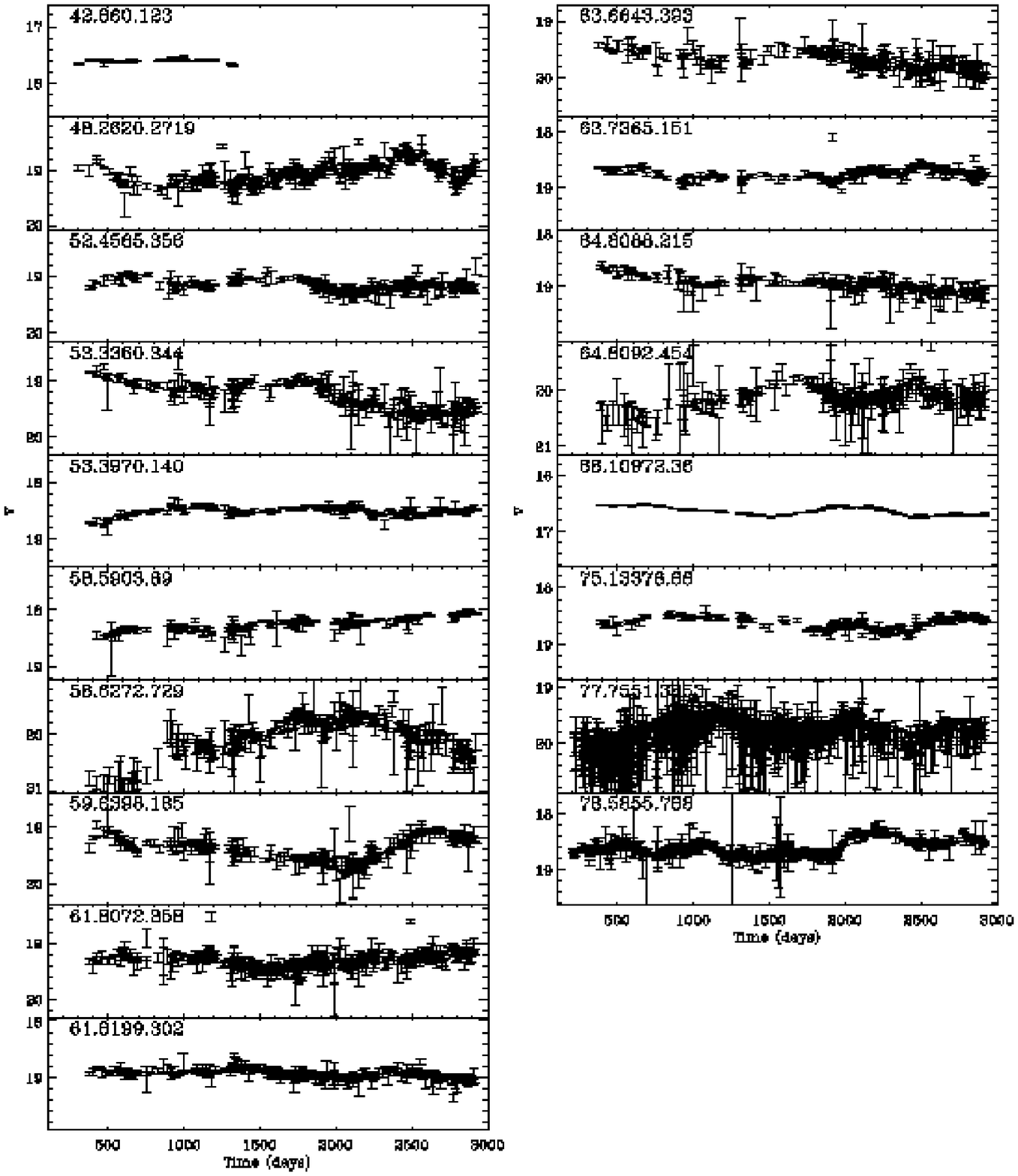}
\vskip 0.5 cm
\addtocounter{figure}{-1}
\caption{continued.}
\end{figure}

\begin{figure}
\epsscale{0.4}
\plotone{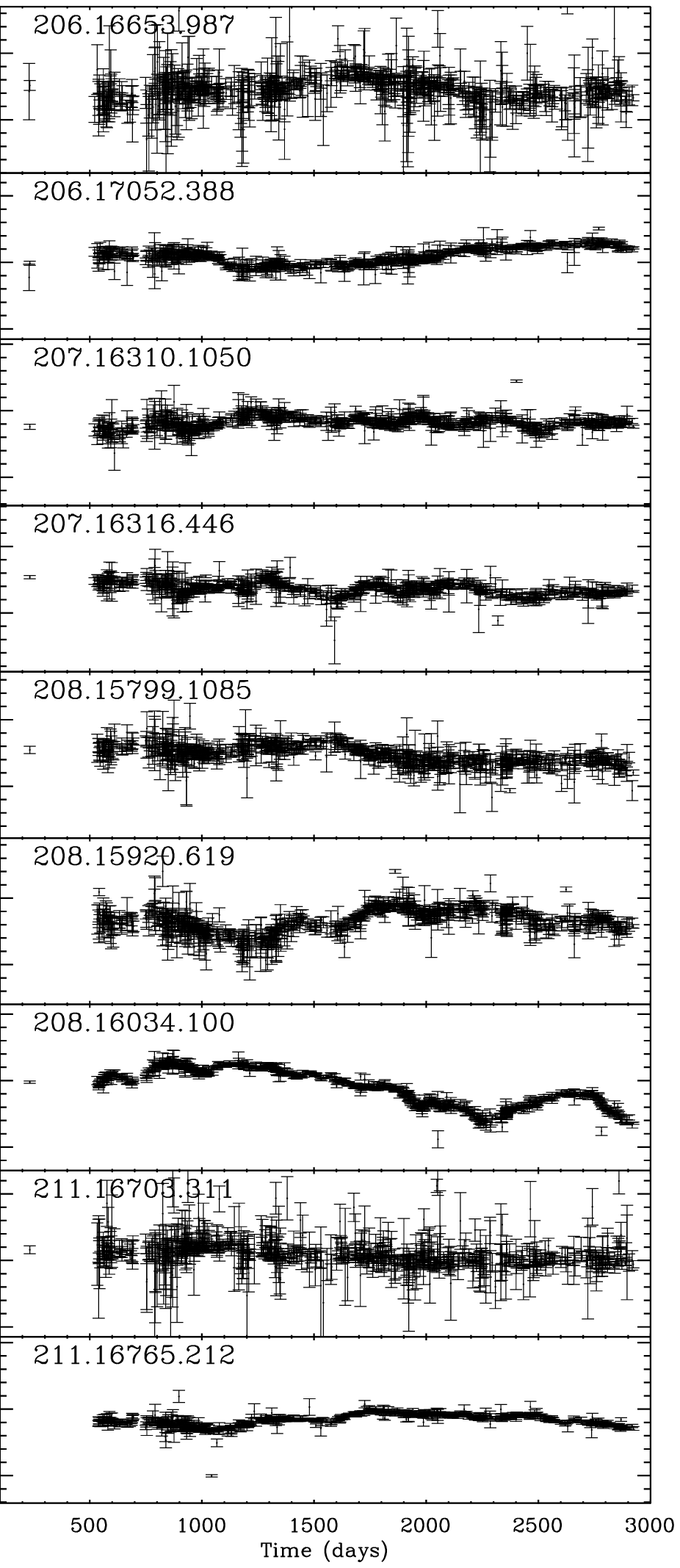}
\vskip 0.5 cm
\caption{$V$-band MACHO lightcurves for 9 quasars discovered behind the
SMC.  See the caption for Figure~\ref{lc}.\label{lc_smc}}
\end{figure}

\clearpage

\begin{deluxetable}{llcccccccc}
\tabletypesize{\scriptsize}
\tablecaption{Previously Identified AGN in the MACHO Database} 
\tablewidth{0pt}
\tablehead{
\colhead{Source Name} &
\colhead{MACHO ID} &
\colhead{$\alpha$ (J2000)} &
\colhead{$\delta$ (J2000)} &
\colhead{$\overline{V}$} & 
\colhead{$(\overline{V-R})$} & 
\colhead{$z$} &
\colhead{$M_V$} & 
\colhead{$n_V$} &
\colhead{Ref}
}
\startdata
050736.52-684751.7 & 1.4418.1930  & 05:07:36.39 & $-$68:47:52.94 & 20.05 & 0.14 & 0.53 & $-$22.59 & 1071 & 1\\
050833.29-685427.5 & 1.4537.1642  & 05:08:31.89 & $-$68:55:10.66 & 19.75 & 0.16 & 0.61 & $-$23.27 & 1137 & 1\\
RX J0509.2-6954    & 5.4643.149   & 05:09:15.49 & $-$69:54:16.75 & 17.95 & 0.28 & 0.17 & $-$21.91 & 939 & 2\\
RX J0524.0-7011    & 6.7059.207   & 05:24:02.31 & $-$70:11:08.95 & 18.26 & 0.42 & 0.15 & $-$21.24 & 980 & 2\\
RX J0517.4-7044    & 13.5962.237  & 05:17:17.03 & $-$70:44:02.46 & 19.33 & 0.40 & 0.17 & $-$20.44 & 894 & 2\\
RX J0531.5-7130    & 14.8249.74   & 05:31:31.60 & $-$71:29:47.78 & 19.36 & 0.23 & 0.22 & $-$21.05 & 869 & 2\\ 
RX J0550.5-7110    & 28.11400.609 & 05:50:31.22 & $-$71:09:58.47 & 20.08 & 0.26 & 0.44 & $-$22.07 & 322 & 2\\
RX J0503.1-6634    & 53.3725.29   & 05:03:04.04 & $-$66:33:46.62 & 18.10 & 0.41 & 0.06 & $-$19.41 & 268 & 2\\
RX J0547.8-6745    & 68.10968.235 & 05:47:45.13 & $-$67:45:5.745 & 20.45 & 0.51 & 0.39 & $-$21.39 & 263 & 2\\
$[$HB89$]$0557-672 & 69.12549.21  & 05:57:22.41 & $-$67:13:22.16 & 17.41 & 0.38 & 0.14 & $-$21.95 & 254 & 3\\
RX J0550.6-6637    & 70.11469.82  & 05:50:33.31 & $-$66:36:52.96 & 18.19 & 0.65 & 0.08 & $-$19.23 & 248 & 2\\
RX J0532-6920      & 82.8403.551  & 05:31:59.66 & $-$69:19:51.12 & 19.40 & 0.31 & 0.15 & $-$20.07 & 851 & 2\\
\enddata
\tablecomments{Weighted average magnitudes $\overline{V}$ and colors
$(\overline{V-R})$ determined from MACHO photometry; $n_V$ is the
number of MACHO photometric data points over the 7.5-year monitoring
period.  Redshifts, $z$,  are taken from discovery papers as follows: 1 =
\citet{dob02}, 2 = \citet{sch99}, 3 = \citet{bla86}.}
\end{deluxetable}

\begin{deluxetable}{lcccccccc}
\tabletypesize{\scriptsize}
\tablecaption{MACHO Quasars Behind the Large Magellanic Cloud} 
\tablewidth{0pt}
\tablehead{
\colhead{MACHO ID} &
\colhead{$\alpha$ (J2000)} &
\colhead{$\delta$ (J2000)} &
\colhead{$\overline{V}$} & 
\colhead{$(\overline{V-R})$} & 
\colhead{$z$} &
\colhead{$M_V$} & 
\colhead{$n_V$} &
\colhead{Notes}
}
\startdata
2.5873.82    & 05:16:28.78 & $-$68:37:02.38 & 17.44 & 0.44 & 0.46 & $-$24.81 & 967 & \\
5.4892.1971  & 05:10:32.32 & $-$69:27:16.90 & 18.45 & 0.33 & 1.58 & $-$26.73 & 960 & \\
6.6572.268   & 05:20:56.93 & $-$70:24:52.50 & 18.33 & 0.24 & 1.81 & $-$27.22 & 989 & \\
9.4641.568   & 05:08:45.95 & $-$70:05:00.92 & 19.20 & 0.30 & 1.18 & $-$25.20 & 993 & \\
9.4882.332   & 05:10:23.18 & $-$70:07:36.12 & 18.83 & 0.32 & 0.32 & $-$22.53 & 1004 & \\
9.5239.505   & 05:12:59.56 & $-$70:30:24.76 & 19.18 & 0.36 & 1.30 & $-$25.48 & 992 & \\
9.5484.258   & 05:14:12.05 & $-$70:20:25.64 & 18.61 & 0.33 & 2.32 & $-$27.37 & 997 & \\
11.8988.1350 & 05:36:00.50 & $-$70:41:28.86 & 19.52 & 0.30 & 0.33 & $-$21.90 & 1008 & a\\
13.5717.178  & 05:15:36.02 & $-$70:54:01.65 & 18.56 & 0.37 & 1.66 & $-$26.74 & 921 & \\
13.6805.324  & 05:22:47.23 & $-$71:01:31.08 & 18.66 & 0.35 & 1.72 & $-$26.39 & 959 & \\
13.6808.521  & 05:22:47.69 & $-$70:47:34.82 & 19.02 & 0.32 & 1.64 & $-$26.26 & 942 & \\
17.2227.488  & 04:53:56.55 & $-$69:40:35.96 & 18.88 & 0.32 & 0.28 & $-$22.13 & 449 & \\
17.3197.1182 & 05:00:17.56 & $-$69:32:16.32 & 18.88 & 0.32 & 0.90 & $-$25.15 & 435 & \\ 
20.4678.600  & 05:08:54.08 & $-$67:37:35.57 & 20.06 & 0.24 & 2.22 & $-$25.80 & 372 & \\
22.4990.462  & 05:11:40.77 & $-$71:00:32.95 & 19.82 & 0.38 & 1.56 & $-$25.32 & 580 & \\
22.5595.1333 & 05:15:22.94 & $-$70:58:06.77 & 18.55 & 0.29 & 1.15 & $-$25.77 & 572 & \\ 
25.3469.117  & 05:01:46.68 & $-$67:32:41.81 & 18.07 & 0.26 & 0.38 & $-$23.72 & 376 & \\
25.3712.72   & 05:02:53.65 & $-$67:25:46.44 & 18.61 & 0.31 & 2.17 & $-$27.20 & 372 & b\\
30.11301.499 & 05:49:41.63 & $-$69:44:15.86 & 19.41 & 0.35 & 0.46 & $-$22.86 & 309 & \\
37.5584.159  & 05:15:04.72 & $-$71:43:38.62 & 19.43 & 0.66 & 0.50 & $-$23.06 & 275 & c\\
42.860.123   & 04:46:11.14 & $-$72:05:09.80 & 17.60 & 0.29 & 0.95 & $-$26.58 & 50 &  d\\
48.2620.2719 & 04:56:14.19 & $-$67:39:10.81 & 19.03 & 0.32 & 0.26 & $-$21.82 & 368 & \\ 
52.4565.356  & 05:08:30.64 & $-$67:02:30.05 & 19.16 & 0.21 & 2.29 & $-$26.78 & 257 & \\
53.3360.344  & 05:00:54.00 & $-$66:44:01.34 & 19.22 & 0.22 & 1.86 & $-$26.39 & 268 & \\
53.3970.140  & 05:04:36.01 & $-$66:24:17.03 & 18.50 & 0.27 & 2.04 & $-$27.14 & 272 & \\
58.5903.69   & 05:16:36.76 & $-$66:34:36.92 & 18.20 & 0.26 & 2.24 & $-$27.68 & 251 & \\
58.6272.729  & 05:18:51.97 & $-$66:09:56.70 & 19.85 & 0.35 & 1.53 & $-$25.24 & 342 & \\
59.6398.185  & 05:19:28.02 & $-$65:49:50.50 & 19.33 & 0.36 & 1.64 & $-$25.95 & 284 & \\ 
61.8072.358  & 05:30:07.93 & $-$67:10:27.20 & 19.33 & 0.27 & 1.65 & $-$25.96 & 388 & \\
61.8199.302  & 05:30:26.81 & $-$66:48:55.31 & 18.94 & 0.25 & 1.79 & $-$26.57 & 392 & \\
63.6643.393  & 05:20:56.45 & $-$65:39:04.79 & 19.65 & 0.41 & 0.47 & $-$22.67 & 250 & \\
63.7365.151  & 05:25:14.29 & $-$65:54:45.93 & 18.72 & 0.33 & 0.65 & $-$24.43 & 252 & \\
64.8088.215  & 05:30:09.06 & $-$66:07:01.05 & 18.96 & 0.23 & 1.95 & $-$26.77 & 257 & \\
64.8092.454  & 05:30:08.75 & $-$65:51:24.27 & 20.10 & 0.20 & 2.03 & $-$25.52 & 258 & \\
68.10972.36  & 05:47:50.18 & $-$67:28:02.44 & 16.63 & 0.28 & 1.01 & $-$27.33 & 267 & \\
75.13376.66  & 06:02:34.25 & $-$68:30:41.51 & 18.63 & 0.26 & 1.07 & $-$25.50 & 241 & e\\
77.7551.3853 & 05:27:16.19 & $-$69:39:33.96 & 19.75 & 0.21 & 0.85 & $-$24.14 & 1471 & f \\
78.5855.788  & 05:16:26.23 & $-$69:48:19.39 & 18.61 & 0.22 & 0.63 & $-$24.48 & 878 & g\\
\enddata
\tablecomments{Weighted average magnitudes $\overline{V}$ and colors
$(\overline{V-R})$ determined from MACHO photometry; $n_V$ is the
number of MACHO photometric data points.  Redshifts, $z$, determined from
spectra discussed in \S\,\ref{spectra}.  The majority of quasars were
selected as candidates based on photometric variability alone.
Variable counterparts to X-ray and radio sources are indicated in the
table notes as follows: (a) RX J0536.0-7041, (b)
1WGA J0508.9-6737 (c) $[$HP99$]$ 1306, (d) PMN J0446-7205, (e) PMN
J0603-6830, (f) 1WGA J0527.2-6939, (g) $[$HP99$]$ 1019.}

\end{deluxetable}

\begin{deluxetable}{lcccccccc}
\tabletypesize{\scriptsize}
\tablecaption{MACHO Quasars Behind the Small Magellanic Cloud} 
\tablewidth{0pt}
\tablehead{
\colhead{MACHO ID} &
\colhead{$\alpha$ (J2000)} &
\colhead{$\delta$ (J2000)} &
\colhead{$\overline{V}$} & 
\colhead{$(\overline{V-R})$} & 
\colhead{$z$} &
\colhead{$M_V$} & 
\colhead{$n_V$} &
\colhead{Notes}
}
\startdata
206.16653.987  & 01:01:27.81 & $-$72:46:14.37 & 19.51 & 0.25 & 1.05 & $-$24.57 & 794 & \\ 
206.17052.388  & 01:07:21.71 & $-$72:48:45.76 & 18.85 & 0.23 & 2.15 & $-$26.93 & 810 & \\
207.16310.1050 & 00:55:59.61 & $-$72:52:45.15 & 19.17 & 0.31 & 1.47 & $-$25.80 & 850 & \\ 
207.16316.446  & 00:55:34.70 & $-$72:28:34.23 & 18.64 & 0.19 & 0.56 & $-$24.17 & 822 & a\\
208.15799.1085 & 00:47:15.76 & $-$72:41:12.24 & 19.52 & 0.26 & 2.77 & $-$26.92 & 861 & \\
208.15920.619  & 00:49:34.43 & $-$72:13:08.99 & 19.28 & 0.18 & 0.91 & $-$24.77 & 858 & \\
208.16034.100  & 00:51:16.89 & $-$72:16:51.06 & 18.03 & 0.25 & 0.49 & $-$24.42 & 878 & \\
211.16703.311  & 01:02:14.36 & $-$73:16:26.80 & 18.92 & 0.34 & 2.18 & $-$26.89 & 791 & \\
211.16765.212  & 01:02:34.73 & $-$72:54:22.20 & 18.15 & 0.29 & 2.13 & $-$27.60 & 795 & \\
\enddata
\tablecomments{See notes for Table 2.  Candidate identified as variable
counterpart to X-ray source: (a) RX J0055.6-7228}
\end{deluxetable}


\begin{thebibliography}{}

\bibitem[Alcalde et~al.(2002)]{alc02} Alcalde, D. et al.\ 2002, \apj, 572, 729

\bibitem[Alcock et~al.(2000)]{alc00} Alcock, C. et al.\ 2000, \apj, 542, 281

\bibitem[Alcock et~al.(1999)]{alc99} Alcock, C. et al.\ 1999, \pasp, 111, 1539

\bibitem[Alcock et~al.(1997)]{alc97} Alcock, C. et al.\ 1997, \apj, 486, 697

\bibitem[Anguita, Loyola \& Pedreros(2000)]{ang00} Anguita, C.,
Loyola, P., Pedreros, M.~H.\ 2000, \aj, 120, 845

\bibitem[Bailey, Glazebrook \& Bridges(2002)]{bai02} Bailey, J.,
Glazebrook, K., \& Bridges, T.\ 2002, 2dF User Manual,
Anglo-Australian Observatory

\bibitem[Blanco \& Heathcote(1986)]{bla86} Blanco, V.~M., \& Heathcote,
S.\ 1986, \pasp, 98, 635


\bibitem[Cook et~al.(1995)]{coo95} Cook, K.~H. et al.\ 1995, in
Astrophysical Applications of Stellar Pulsation, ASP Conf.~Series 83,
ed. R.~S.~Stobies \& P.~A.~Whitelock (San Fransisco:ASP), 221

\bibitem[Crampton et~al.(1997)]{cra97} Crampton, D., Gussie, G., Cowley, A.P., \&
Schmidtke, P. C., \aj, 114, 2353

\bibitem[Cristiani et~al.(1997)]{cri97} Cristiani, S., Trentini, S., La
Franca, F., \& Andreani, P.\ 1997, A\&A, 321, 123

\bibitem[Dobrzycki et~al.(2002)]{dob02} Dobrzycki, A., Groot, P.~J.,
Macri, L.~M., \& Stanek, K.~Z.\ 2002, \apjl, in press
(astro-ph/0202524)

\bibitem[Dutra et~al.(2001)]{dut01} Dutra, C.~M., Bica, E., Claria,
J.~J., Piatti, A.~E., \& Ahumada, A.~V.\ 2001, A\&A, 371, 895

\bibitem[Eyer(2002)]{eye02} Eyer, L.\ 2002, AcA, submitted
(astro-ph/0206074)

\bibitem[Filipovic et~al.(1998)]{fil98} Filipovic, M.~D., Haynes,
R.~F., White, G.~L., \& Jones, P.~A.\ 1998, A\&AS, 130, 421

\bibitem[Filipovic et~al.(1997)]{fil97} Filipovic, M.~D., Jones,
P.~A., White, G.~L., Haynes, R.~F., Klein, U., \& Wielebinski, R.\ 1997,
A\&AS, 121, 321

\bibitem[Gibson et~al.(2000)]{gib00} Gibson, B.~K., Giroux, M.~L.,
Penton, S.~V., Putman, M.~E., Stocke, J.~T., Shull, J.~M.\ 2000, \aj,
121,922

\bibitem[Giveon et~al.(1999)]{giv99} Giveon, U., Maoz, D., Kaspi, S.,
Netzer, H., \& Smith, P.~S.\ 1999, \mnras, 306, 637

\bibitem[Haberl et~al(2001)]{hab01} Haberl, F., Dennerl, K.,
Filipovic, M.~D., Aschenbach, B., Pietsh, W., \& Tr\"{u}mper, J.\
2001, A\&A, 365, L208

\bibitem[Haberl \& Pietsch(1999)]{hab99} Haberl, F., Pietsch, W.\ 1999,
A\&AS, 139, 277

\bibitem[Hawkings (2002)]{haw02} Hawkings, M.~R.~S.\ 2002, \mnras, 329, 76

\bibitem[Hewett et~al.(1995)]{hew95} Hewett, P.~C., Foltz, C.~B., \&
Chaffee, F.~H.\ 1995, \aj, 109, 1499

\bibitem[Hjorth et~al.(2002)]{hjo02} Hjorth, J. et al.\ 2002, \apj, 572, L11

\bibitem[Hook et~al.(1994)]{hoo94} Hook, I.~M., McMahon, R.~G, Boyle,
B.~J., \& Irwin, M.~J.\ 1994, \mnras, 268, 305

\bibitem[Hubert \& Floquet(1998)]{hub98} Hubert, A.~M., \& Floquet, M.\ 1998, A\&A, 335, 565

\bibitem[Jones, Klemola \& Lin(1994)]{jon94} Jones, B.~F., Klemola,
A.~R., Lin, D.~N.~C.\ 1994, \aj, 107, 1333

\bibitem[Kahabka et~al.(2001)]{kah01} Kahabka, P., de Boer, K.~S., \&
Br\"{u}ns, C.\ 2001, A\&A, 371, 816

\bibitem[Kahabka et~al.(1999)]{kah99} Kahabka, P., Pietsch, W.,
Filipovic, M.~D., \& Haberl, F.\ 1999, A\&AS, 136, 81


\bibitem[Keller et~al.(2002)]{kel02} Keller, S., Bessell, M.~S., Cook,
K.~H., Geha, M., \& Syphers, D.\ 2002, \aj, accepted (astro-ph/0206444) 

\bibitem[Kroupa \& Bastian(1997)]{kro97} Kroupa, P., Bastian, U.\ 1997, New Astronomy, 2, 77

\bibitem[Marx, Dickey \& Mebold(1997)]{mar97} Marx, M., Dickey, J.~M.,
\& Mebold, U.\ 1997, A\&AS, 126, 325

\bibitem[Meusinger \& Brunzendorf(2002)]{meu02} Meusinger, H., \&
Brunzendorf, J.\ 2002, A\&A, 374, 878

\bibitem[Prochaska, Ryan-Weber \& Staveley-Smith(2002)]{pro02}
Prochaska, J.~X., Ryan-Weber, E., Staveley-Smith, L.\ 2002, \pasp,
accepted (astro-ph/0207479)

\bibitem[Sasaki, Haberl \& Pietsch(2000)]{sas00} Sasaki, M., Haberl,
F., \& Pietsch, W.\ 2000, A\&AS, 143, 391

\bibitem[Schmidtke et~al.(1999)]{sch99} Schmidtke, P. C., Cowley, A.,
Crane, J., Taylor, V., McGrath, T., Hutchings, J., \& Crampton, D.\
1999, \aj, 117, 927

\bibitem[Sirola et~al.(1998)]{sir98} Sirola, C.~J.\ 1998, \apj, 495, 659

\bibitem[Tinney(1999)]{tin99} Tinney, C. G.\ 1999, \mnras, 303, 565


\bibitem[Welch \& Stetson(1993)]{wel93} Welch, D.~L., Stetson, P.~B.\
1993, \aj, 105, 1813

\end{thebibliography}
\end{document}